\documentclass[pre,floatfix,showpacs,superscriptaddress,preprint]{revtex4}
\usepackage{latexsym}
\usepackage{amstext,amsfonts}
\usepackage{epsfig}
\usepackage{url}
\usepackage[usenames]{color}

\begin{document}

\title{Using relaxational dynamics to reduce network congestion}
\author{Ana L. Pastore y Piontti}
\affiliation{Departamento de F\'{\i}sica, Facultad de Ciencias
Exactas y Naturales, Universidad Nacional de Mar del Plata, Funes
3350, 7600 Mar del Plata, Argentina}
\author{Cristian E. La Rocca}
\affiliation{Departamento de F\'{\i}sica, Facultad de Ciencias
Exactas y Naturales, Universidad Nacional de Mar del Plata, Funes
3350, 7600 Mar del Plata, Argentina}
\author{Zolt\'an Toroczkai}
\affiliation{Department of Physics and Center for Complex Network
Research, University of Notre Dame, 225 Nieuwland Science Hall,
Notre Dame, IN, 46556} \affiliation{Theoretical Division, Los
Alamos National Laboratory, Mail Stop B258, Los Alamos, NM 87545
USA}
\author{Lidia A. Braunstein}
\affiliation{Departamento de F\'{\i}sica, Facultad de Ciencias
Exactas y Naturales, Universidad Nacional de Mar del Plata, Funes
3350, 7600 Mar del Plata, Argentina}
\affiliation{Center for polymer studies, Boston University, Boston, MA 02215, USA}
\author{Pablo A. Macri}
\affiliation{Departamento de F\'{\i}sica, Facultad de Ciencias Exactas y
Naturales, Universidad Nacional de Mar del Plata, Funes 3350, 7600 
Mar del Plata, Argentina}
\author{Eduardo L\'{o}pez}
\affiliation{Theoretical Division, Los Alamos National Laboratory,
Mail Stop B258, Los Alamos, NM 87545 USA}

\date{\today}

\begin{abstract}
We study the effects of relaxational dynamics on congestion
pressure in scale free networks by analyzing the properties of the
corresponding gradient networks \cite{Znature}. Using the Family
model \cite{Family} from surface-growth physics as single-step
load-balancing dynamics, we show that the congestion pressure
considerably drops on scale-free networks when compared with the
same dynamics on random graphs. This is due to a 
structural transition of the corresponding gradient network
clusters, which self-organize such as to reduce the congestion
pressure. This reduction is enhanced when lowering 
the value of the connectivity exponent $\lambda$ towards $2$.

\end{abstract}
\maketitle

Transport networks, such as computer networks (Internet), airways, energy
transportation networks, etc., are amongst the most vital components of
modern day infrastructures.  These large-scale networks have not been
globally designed, instead they are the result of local processes. It has
been observed that many networks have a scale-free (SF) connectivity
structure \cite{Reviews, InternetRev}.  Scale-free networks are characterized
by a power-law degree distribution $P(k) \sim k ^{-\lambda}$, ($k \ge k_{min}
$), where $\lambda$ is the connectivity exponent and $k_{min}$ is the lower
degree that a node can have. There have been a number of mechanisms proposed
in the form of stochastic network growth models that produce SF structures
\cite{Reviews, InternetRev} including weighted versions of these processes
\cite{BBV04} (for discussions on the utility of SF models see
Ref. \cite{critics}). However, these models do not {\em explicitly} connect
the flow dynamics and transport performance (such as throughput, queuing
characteristics, etc.) with network topology.  This is difficult to do in
general, since the time-scales of the flow on the network and that of the
network's structural evolution itself can be rather different.

\begin{figure}[htbp]
\vspace*{6mm} \epsfig{file=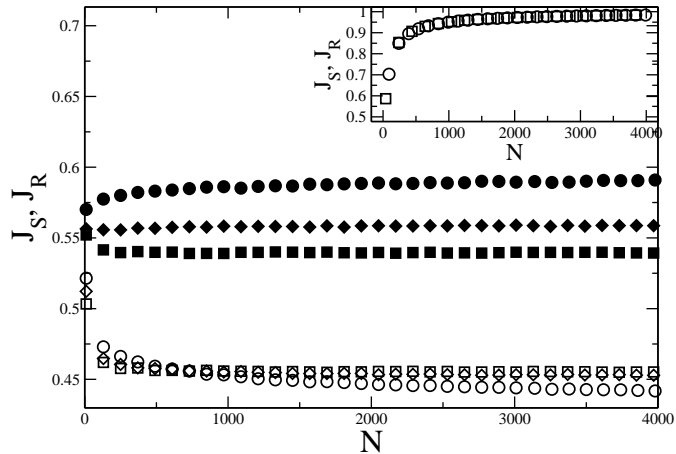,width=3.0in,angle=-90} \caption{
$J_{S}$ and $J_{R}$ as function of $N$ for SF networks with
different values of $\lambda$, $2.5$ ($\bigcirc$), $3$($\diamond$),
$3.5$ ($\square$), with filled symbols for S and empty symbols for
R. In the inset we plot $J_{S}$, ($\bigcirc$), and $J_{R}$,
($\square$), as function of $N$ for ER networks ($p=0.1$).
\label{Jamayd}}
\end{figure}

\begin{figure}[htbp]
\vspace*{0mm} \epsfig{file=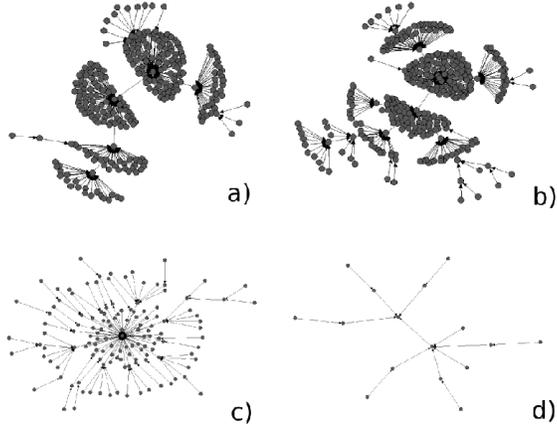,width=3.0in} \caption{Largest
clusters of the gradient networks of ER and SF with $\lambda=2.5$
and $N=1024$. In the top plots,  we show the structure for the ER
and on the bottom  for SF. The left plots (a),c)) correspond to the S and
the right ones (b), d)) to the R models.\label{esq}}
\end{figure}

In this Letter we show that the emergence of scale-free structures
is favored against non-scale-free structures, such as random
graphs, if the transport dynamics has a relaxation component
(called load-balancing in communications). We will see that {\em
even one step} of such a gradient flow \cite{Znature} will
considerably reduce the congestion pressure in scale-free networks
while it has virtually {\em no effect} in random graphs. In addition, within the class
of uncorrelated scale-free networks, the congestion reduction is enhanced for
low (close to 2) $\lambda$ values. Although we use
the jargon from the fields of communication networks and queuing
theory, we expect that our results hold for large-scale
networks in general, where the flow dynamics is induced at least
in part by the existence of gradients, a rather ubiquitous
mechanism. In the following, by ``packet'' we mean any discrete
entity transported between two nodes of a network of $N$ nodes.
We assume that the network is driven in  ``the volume'', by
packets entering at random at an average rate $\gamma$ at any of
the nodes (this is realistic since the users actions in general
are uncorrelated) \cite{ohira,Fuks99,Arenas01,Guimera02}. Using
the language of queuing theory, if a node in the time interval
$(t,t+\tau)$ sends packets into the network, but it receives no
packets from any of its neighbors, we say that it acts as a
``client'', while if it receives a packet or several more from its
neighbors, we say that it acts as a ``server'' \cite{explain}.
Here $\tau > 0$ is the average processing time of a single packet
by a node.  Measuring the average fraction $J  = \langle
N_c/N\rangle$ of the number of clients $N_c$ (over a period of
time in the steady state), gives us a simple global measure for
the congestion {\em pressure} present in the network
\cite{Znature}.  The average $ \langle \cdot \rangle$ is over the
randomness in the input but it can also be over network structure
when comparing {\em classes} of networks. The client nodes are the
ones that introduce new packets into the network, but they do not
contribute to routing. Obviously, higher $J$ means more
congestion. Certainly, all networks will become congested at large
enough driving rates $\gamma > \gamma_c$
\cite{ohira,Fuks99,Arenas01,Guimera02, Ech04, Danila}. $J$
indicates which network will become congested earlier, larger $J$
meaning smaller $\gamma_c$.  $J$ is a global indicator that, however, does
not take into account the {\em distribution} of the packets over
the server nodes.  That can be done via betweenness-based
quantities as in Ref. \cite{Sameet}.
Load-balancing is a specific case of the more general process of
gradients induced flows \cite{Znature},  where the flows are
produced by the local gradients of a non-degenerate scalar field
${\bf h} = \{h_i\}_{i=1}^{N}$ distributed over the $N$ nodes of a
substrate graph $G$ (transport network). The scalar field could
represent, for example, the number of packets at the routers
\cite{Rabani98,Valverde}, or the virtual time horizon of the
processors in parallel discrete event simulations \cite{Korniss}.
The gradient direction of a node $i$ is a directed edge pointing
towards that neighbor (on $G$) $j$ of $i$ which has the lowest
value of the scalar in $i$-s neighborhood. If $i$ has the lowest
value of $h$ in its network neighborhood, the gradient link is a
self-loop. The gradient network $\nabla_h G$ is defined simply as
the collection of all gradient edges on the substrate graph $G$
 \cite{Znature,Toroczkai-cond-mat}.
It represents the subgraph of {\em instantaneous maximum flow} if
the flow is induced by these gradients. 
In the gradient network each node has a unique
outgoing link and $\ell$ incoming links. When a node has $\ell=0$,
(no inflow in that instant), it acts as a client, otherwise it is
a server. Then certainly, $J$ is the average fraction of nodes,
with $\ell=0$, i.e., it is the fraction of the ``leaves'' of
$\nabla_h G$. Note that $J$ is a queuing characteristic, rather
than an actual throughput measure.  It was shown \cite{Znature,Toroczkai-cond-mat}
that distributing random scalars $\{h\}$ independently onto the
nodes of a network $G$, to which we refer as the {\em static} (S)
model in the remainder, Erd\H{os}-R\'enyi (ER) graphs \cite{ER}
(with given link probability $p$) become more congested with
increasing network size $N$, i.e., $J_{S}\rightarrow 1$ while on
SF networks $J_{S}$ converges to a finite sub-unitary value, see
the plots for $J_S$ in Fig. \ref{Jamayd}.
Once the (gradient) flow commences through the network, the scalar
field becomes correlated and the queuing characteristics change.
Usually, packets have destinations, and thus they cannot be
governed exclusively by gradient flows, however, relaxational
dynamics can be employed for finite periods of time. In the
following we systematically study the effects of a {\em single}
relaxational step, and show, that even in this case, the effects
on congestion pressure can be drastic. First, we note that the
one-step relaxation dynamics defined by the gradient flow is 
nothing but the deposition model with surface relaxation (Family model)
\cite{Family} from surface-growth physics extended to networks
\cite{Pastore}. To generate SF networks, we used the
configurational model \cite{ConfigModel} with $k_{min}=2$
\cite{explainkmin} (mostly for mathematical convenience, but 
see the discussion in the end about other networks). 
 At $t=0$ a random scalar field $h$ is
constructed by assigning to each node of the substrate network a
random scalar independently and uniformly distributed between $0$
and $1$.  At this stage the initial static gradient network
\cite{Znature} is formed and its jamming coefficient $J_{S}$
determined. Then the scalars $h\equiv h(t)$ are evolved obeying
the rules of the Family model \cite{Family}: at every time step a
node $i$ of the substrate is chosen at random with probability
$1/N$ and it becomes a candidate for growth. If $h_i < h_j$ for
every $j$ (gradient criterion) which is a nearest neighbor of the
node $i$, $h_i\rightarrow h_i+1$. Otherwise, if $h_i$ is not a
minimum, the node $j$ with minimum $h$ is incremented by one. When
the process reaches the steady state \cite{Pastore} of the
evolution with this relaxation (R), we construct the gradient network
and measure $J_{R}$. In accordance with previous observations
\cite{ZTandKorniss,Pastore}, the steady state is reached extremely
fast: the saturation time actually does not scale with the system
size $N$ but it approaches an $N$-independent constant.
\begin{figure}[htbp]
\vspace*{5mm} \epsfig{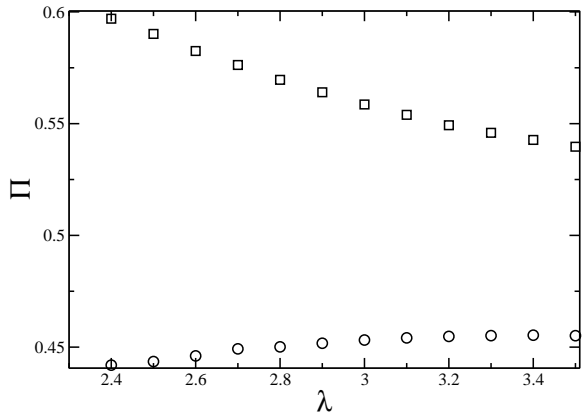} \caption{Plot of
$\Pi$, $\bigcirc$ for R and $\square$ for S,  for SF networks as
function of $\lambda$ ($N=4000$).  \label{pi}}
\end{figure}
In Fig. \ref{Jamayd} we plot $J_{S}$ and $J_{R}$ as function of $N$ for ER
graphs and for SF networks (for the latter we compare cases with different
$\lambda$ values). As one can observe (inset), for ER networks, the model
with relaxation has no effect on lowering the congestion, i.e., $J_{S} \simeq
J_{R}$.  For SF networks, however, there is a drastic difference between the
static and dynamic cases, with $J_{S}$ being considerably larger than $J_{R}$
for large enough $N$.  Note that $J_S$ increases with decreasing $\lambda$,
which can be understood through the fact that for lower $\lambda$ values the
$\nabla_h G$ of the SF graph is increasingly star-like, creating more
congestion (see below).  Since in real-world networks, however, one expects
to find a load-balancing component of the transmission dynamics (see
\cite{Rabani98}), our model with relaxation is a better representation than
the static one.  And indeed, from Fig.  \ref{Jamayd} it becomes apparent that
$J_R$ has the {\em opposite} behavior as function of $\lambda$ for large
enough networks: lowering $\lambda$ lowers the congestion pressure $J_R$.
Next we show how the drop in the congestion pressure due to relaxation
dynamics can be understood in terms of a structural change of the clusters of
the corresponding gradient network. Here the clusters are defined as the
disconnected components (trees) of the gradient graph. The decrease of
$J_{R}$ (which is the fraction of leaves of the gradient network) means a
decrease of the perimeter of the clusters of $\nabla_h G$.  To simplify the
discussion, in the following we will use the symbol $\Pi$ to denote the
fraction of leaves (or the perimeter) of $\nabla_h G$, and thus $\Pi = J_{S\;
or R}$. In Fig.~\ref{pi} we show $\Pi$ as function of $\lambda$ for a fixed
network size $N$.  From this we can see that for a given value of $N$, $\Pi$
is larger for the S model than the R one.  This is compatible with a
transition on the structure of the trees of the gradient network from a
star-like structure in the S state to a more elongated structure in the R,
see Fig.~\ref{esq}. This transition is responsible for the drop in the
congestion pressure after the relaxation step is applied. From Fig.~\ref{esq}
we can see that for the ER networks the R model does not affect significantly
the structure of the clusters of the gradient network (going from Fig.
~\ref{esq}a) to b)), meanwhile, for SF networks (going from Fig.
~\ref{esq}c) to d)) we observe a major structural transition before and after
the relaxation dynamics is applied, from a star-like cluster to an elongated
one.
For a quantitative insight {\em of the S model} consider that
the scalars are identically (and independently) distributed according to some
distribution $\eta(h)$.  For the calculations below we assume that the SF
network has low clustering, i.e., that the probability of two neighbors of a
node to be also neighbors is small.  
Assume that the central node $i$ has its scalar value equal to $h_i$.
Then, the  probability that the
{\em neighbor} $j$ of $i$ points its gradient link (given $h_i$) into $i$ equals to:
\begin{equation}
p_{j\to i} \Big|_{h_i} =  \left[ \int_{h_i}^{1} dh'\;\eta(h') \right]^{m_j} \equiv 
[\phi(h_i)]^{m_j}\;,  \label{pjih}
\end{equation}
where $\phi(x) = \int_{x}^1 dh \;\eta(h)$ and $m_j$ is the degree of node $j$. 
This is because the scalars at
all the $m_j-1$ neighbors of $j$ (that do not include $i$) must be larger
than  $h_i$, and in addition we also have to have $h_j >
h_i$. Hence the probability that node $j$ {\em does not} point its gradient link into
$i$ is $(1-p_{j\to i} \big|_{h_i})$.
In order for node $i$ to be a leaf on the gradient tree, one must have that none
of its neighbors point their gradient directions into it. 
This is given by: $\prod_{j=1}^{k_i} (1-p_{j\to i} \big|_{h_i})$, where
for simplicity we labelled the neighbors of $i$ by $j=1,2,\ldots,k_i$.
Thus, the probability that $i$ is a leaf  is expressed by:
$\pi_i = \int_0^1 dh_i \;\eta(h_i) \prod_{j=1}^{k_i}(1-p_{j\to i} \big|_{h_i})$.
Using (\ref{pjih}) and 
the change of variable from $h$ to $\phi$, $d\phi = - dh\;\eta(h)$, the integral becomes:
\begin{equation}
\pi_i = \pi_i(m_1,m_2,\ldots,m_{k_i})=\int_0^1 d\phi \prod_{j=1}^{k_i} \left(1-\phi^{m_j} \right)\;. \label{exact}
\end{equation}
\begin{figure}[htbp]
\vspace*{4mm} \epsfig{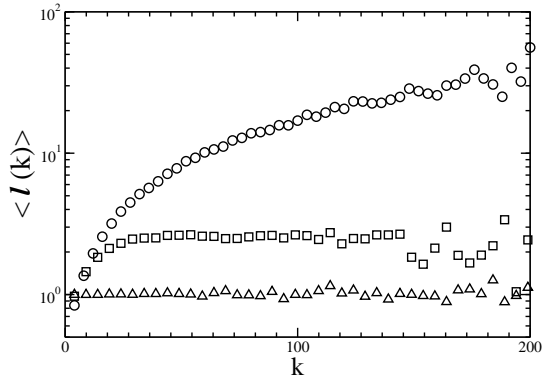}
\caption{Linear-$\log$ plot of $\langle \ell(k) \rangle$ as
function of $k$
 with $\lambda=2.5$ and $N=1024$ for S ($\bigcirc$),  and R
  ($\square$), for $t=0.1$ and  R ($\bigtriangleup$) in the steady state. As the
  system evolves, $\langle \ell(k) \rangle$ loses its dependence on $k$. We show
  the plot in the lin-$\log$ scale in order to be able to visualize them in
  the same figure.\label{lk}}
\end{figure}
This expression shows that among all nodes $i$ with the same degree $k$,
those will likely be leaves in $\nabla_h G$, which have neighbors with {\em
high degree}. It also shows that hubs will have very low probability of
becoming leaves since in that case many ($k$) sub-unitary values are
multiplied in Eq.~(\ref{exact}). Therefore, leaves are coming from the set of
nodes with low, or moderate degrees, connected to hubs, supporting
Fig.~\ref{esq}c).
\begin{figure}[htbp]
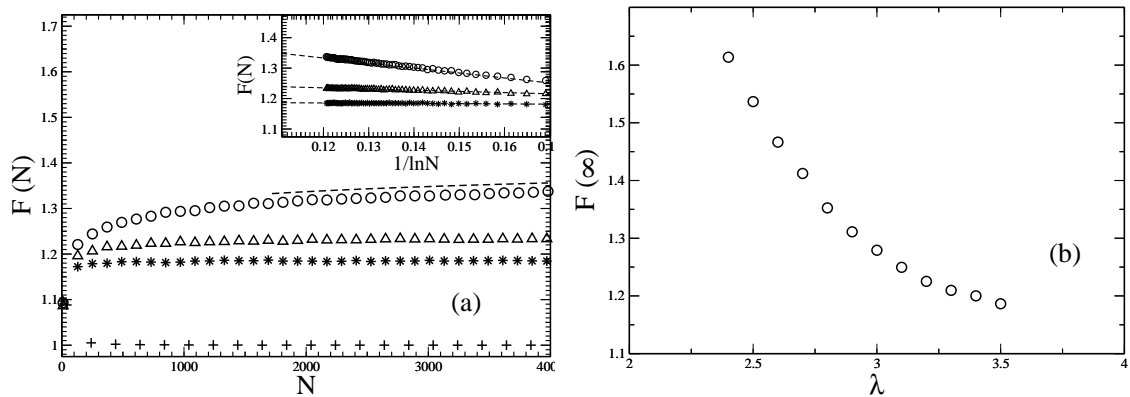

\vspace*{6mm} \epsfig{file=fig5a.eps,width=2.9in} 
\epsfig{file=fig5b.eps,width=2.9in,angle=0}
\caption{  a) $F(N)$ as function of $N$, for SF networks with
$\lambda=2.5$ ($\bigcirc$), $3.0$ ($\bigtriangleup$), $3.5$ ($*$) and for ER
networks with $p=0.1$ ($+$). The dashed lines correspond to the logarithmic
fit of $F(N)$ as function of $N$ in the asymptotic regime (see also
inset). b) Asymptotic improvement factor as function of the connectivity
exponent.\label{J_RyJ_S}}
\end{figure}
When there is relaxation, the leaves are sending their load towards higher
degree nodes (on average), correlating the scalars with the degree. This
increases the chance of larger degree nodes to become leaves and breaking the
star-like structures. As a consequence, the gradient network makes a
transition to more elongated clusters, reducing its perimeter, and
thus the congestion pressure. To support our claims numerically, we computed
the average in-degree $\langle \ell(k) \rangle$ of $\nabla_h G$ as function
of $k$ in both the S and the R models. In the S model the gradient clusters
are star-like, and thus they are very heterogeneous, which appears as a
strong dependence of $\langle \ell(k) \rangle$ on $k$, see the circles in
Fig.~\ref{lk}. After relaxation, however, this heterogeneity almost
disappears (squares and triangles in Fig.~\ref{lk}), and in the steady state,
$\langle \ell(k) \rangle \approx 1$ (only a weak $k$-dependence remains),
which means that $\nabla_h G$ is made of elongated, homogenous trees.
Defining the relative improvement factor as $F(N)=J_{S}/J_{R}$, which
measures how efficient is the relaxation dynamics in reducing the congestion
\cite{nota}, we find that for SF networks, as $\lambda$ decreases, $F(N)$
increases, see Fig.~\ref{J_RyJ_S}a), where we plot $F(N)$ as function of $N$
for different values of $\lambda$.  In addition, from Fig.~\ref{J_RyJ_S}a),
we observe that $F(N)$ has a logarithmic convergence to its asymptotic value,
i.e., $F(N) \simeq F(\infty) - K/\ln N$ (with $K$ as a constant), see also
the inset of Fig.~\ref{J_RyJ_S}a). Fig.~\ref{J_RyJ_S}b) shows the
infinite-system size relative improvement factor as function of $\lambda$,
showing the increasing power of the relaxation mechanism for lower $\lambda$
values.  The effects shown here seem to hold for other SF networks that we
tested as well, including real-world networks such as the DIMES Internet
mapping project generated AS level network (www.netdimes.org). The DIMES
network is a correlated SF graph, and there the relaxation dynamics improves
the congestion pressure drastically: from a $J_S=0.61$ the congestion
pressure drops to $J_R = 0.18$! This suggests that topology correlations play
an important role which are the subject of present studies. \\
\noindent {\em Acknowledgments.} This work was supported by UNMdP
and FONCyT (PICT 2005/32353). P. A. M. is also a member of
CONICET. Z.T., L.A.B. and E.L. were supported in part by the US
DOE grant W-7405-ENG-36.


\begin{thebibliography}{99}
\bibitem{explain} Packet flow is a continuous-time discrete event  process.
However, in the non-infinitesimal time $\tau > 0$ of processing a single packet, several packets may enter a node's
queue from its neighbors or from being generated at that node.
\bibitem{explainkmin} For $k_{min} > 1$ there is a high probability that the
network is fully connected, wich is required in our model see also
\cite{Pastore}.
\bibitem{nota} Notice that $\cal{F}$$>1$ means that the congestion is reduced.
\end{thebibliography}
\end{document}